\documentclass[aps,floatfix,showpacs,amsmath,superscriptaddress,nofootinbib,
12pt,a4paper]{revtex4-1}
\usepackage{graphicx}

\begin{document}

\preprint{IZTECH-P2011-03}

\title{A simple toy model for a unified picture of dark energy, 
dark matter, and inflation}

\author{Recai Erdem}
\email{recaierdem@iyte.edu.tr}
\affiliation{Department of Physics,
{\.{I}}zmir Institute of Technology \\ 
G{\"{u}}lbah{\c{c}}e K{\"{o}}y{\"{u}}, Urla, {\.{I}}zmir 35430, 
Turkey} 

\date{\today}

\begin{abstract}
A specific scale factor in Robertson-Walker metric with the prospect of 
giving the overall cosmic history in a unified picture roughly is 
considered. 
The corresponding energy-momentum tensor is identified as that of two scalar 
fields where one plays the roles of both inflaton and dark matter while 
the other accounts for dark energy. A preliminary phenomenological 
analysis 
gives an order of magnitude agreement with observational data. The 
resulting picture may be considered as a first step towards a single 
model for all epochs of cosmic evolution. 
\end{abstract}

\maketitle

\section{introduction}
There is an intense on-going research to understand the natures of 
late-time acceleration \cite{PDG} (whose standard explanation is 
dark-energy \cite{dark-energy}), dark-matter \cite{dark-matter}, and the 
inflationary era 
\cite{inflation1,inflation2}. A detailed and definite formulation 
of each of these issues by its own is essential and very important for the 
future direction of cosmology. However how to relate these in a 
single formulation and unify ally eras of cosmology (namely, inflationary, 
radiation dominated, matter dominated, the current late-time acceleration) 
is as essential as the study of each era separately. This is not only due 
to the fact that we must eventually put all these into single picture but it 
is necessary for a better and correct formulation each of these issues. 
This paper is an attempt in this direction i.e. to obtain an overall 
picture of cosmic history in a single model. Some of the other studies in 
this direction may be found in \cite{Guendelman,Sola,others}. I hope the 
study given here is simpler and more concrete while being minimal and 
formulated in standard framework i.e. two standard scalar fields in the 
usual 4-dimensional Robertson Walker metric and the usual Einstein-Hilbert 
gravity. 

In this paper I consider a specific scale factor in the usual 
4-dimensional Robertson-Walker metric. The scale factor is chosen in a 
such a way that it has a prospect to account for inflationary, matter 
dominated, and late-time acceleration eras of the universe.  Then I check 
this expectation. First I find the corresponding energy-momentum tensor 
and identify it by that of two scalar fields. The first one mimics 
inflation at very small times and then mimics (dark) matter at 
intermediate times. The second one is identified by dark energy. Hence 
this model accounts for all epochs of the universe except the radiation 
dominated one. The content of this universe 
is similar to our own except it does not contain baryonic matter and 
radiation. This universe is similar to our own, given the fact that the 
present ratio of baryonic matter density and radiation to the total energy 
density of the universe is 4 $\%$ and hence negligible and remains 
negligible (at gravitational level) in most 
of the cosmological evolution except in the radiation dominated era.
Then I use cosmological data to constraint the 
parameters of the model and apply these to some 
redshifts and to the corresponding time data to check the 
phenomenological viability of the model. There is an order of magnitude 
agreement with data. In my opinion the results are encouraging to look  
for a more elaborate form of the model where baryonic matter and radiation 
are included, and where a more thorough study of the parameter space 
is investigated.

\section{the model}
We consider the Robertson-Walker metric
\begin{equation}
ds^2\;=\;g_{\mu\nu}dx^\mu\,dx^\nu\,=\,-dt^2\,+\,a^2(t)\tilde{g}_{ij}dx^idx^j
\label{a1} 
\end{equation}
We take the 3-dimensional space be flat, i.e. 
$\tilde{g}_{ij}=\delta_{ij}$ for the sake of simplicity, which is an 
assumption consistent with cosmological observations \cite{PDG}. The key 
assumption in this paper is the following ansatz
\begin{equation}
a(t)\,=\,[p_1\,+\,p_2a_2t]\,Exp[-b_1(a_2t)^{-1/6}]
\label{a2} 
\end{equation}
where $p_1$, $p_2$, $a_2$, $b_1$ are some constants that to be fixed or 
bounded by consistency arguments or cosmological observations. The 
corresponding Hubble constant and its rate of change are given by
\begin{eqnarray}
&&H\,=\,\frac{\dot{a}}{a}=\frac{a_2[6p_2(a_2t)^{\frac{7}{6}}+b_1(p_1+p_2a_2t)]}
{6(a_2t)^{\frac{7}{6}}(p_1+p_2a_2t)} \label{a3a} \\
&&\frac{\dot{H}}{H^2}
\,=\,-\frac{(a_2t)^{\frac{1}{6}}
[36p_2^2(a_2t)^{\frac{13}{6}}+7b_1(p_1+p_2a_2t)^2]}
{[6p_2(a_2t)^{\frac{7}{6}}+b_1(p_1+p_2a_2t)]^2}
\label{a3b}
\end{eqnarray}
and the acceleration of the scale factor is
\begin{equation}
\frac{\ddot{a}}{a}\,=\,\frac{b_1[b_1(p_1+p_2a_2t)
+(a_2t)^{\frac{1}{6}}(-7p_1+5p_2a_2t)]}{36t^2(a_2t)^{\frac{1}{3}}(p_1+p_2a_2t)}
\label{a4}
\end{equation}
where the dots on top of the letters stand for time derivative. 

The 
following observations about the scale factor $a(t)$ are in order; One 
notices that $a(t)$ is positive for all values of t provided that
\begin{equation}
p_1,p_2,a_2,b_1\,>\,0 \label{a5}
\end{equation}
$\frac{\ddot{a}}{a}$ is positive for 
extremely small values of $a_2t$, where the leading term in 
$\frac{\ddot{a}}{a}$ is the $p_1$ term; $\frac{\ddot{a}}{a}$ is negative 
for the intermediate values of $a_2t$, where the leading term in the 
numerator is the $-7(a_2t)^{\frac{1}{6}}$ term;  and 
$\frac{\ddot{a}}{a}$ is positive again for 
the larger values of $a_2t$. Note that the present era corresponds to 
very large values of $t$, not the infinite value of $t$ where the 
acceleration is zero. One may see the general form of the evolution 
of $\frac{\ddot{a}}{a}$ for a set of phenomenologically relevant 
parameters in the next section in Figure \ref{fig:acc}. 
Moreover it is evident from Eq.(\ref{a3b}) 
that $\frac{\dot{H}}{H^2}$ here is almost zero (i.e. slow-condition is 
satisfied) if $a_2t$ is taken sufficiently small. 
Therefore the scale factor ansatz given above, at least in 
principle, is suitable to account for all four eras  of cosmic expansion; 
inflation, radiation dominated  era, matter dominated era, and current 
accelerated expansion era. In the analysis given below first I will 
determine the Einstein tensor and the corresponding energy-momentum 
tensor. I will identify this energy-momentum tensor with that of two 
scalar fields. Then, after using phenomenological 
considerations, the parameters (i.e. $p_1$, $p_2$, $a_2$, $b_1$) are  
numerically constrained. I will check the 
phenomenological viability of the model. It will be seen that the scalar 
fields may be identified by inflaton, dark energy and dark matter, and the 
corresponding picture is that of a universe that consists of only dark 
energy and dark matter (that also serves as inflaton at early times). 
Given the fact that, at present, more than 96 $\%$ of the universe consist 
of dark energy and dark matter this universe will be 
considered as a universe that is similar to our own in its 
overall cosmic history except in the radiation dominated era. Although the 
results obtained here have only order 
of magnitude agreement with observations, the results are encouraging for 
adopting this model as a starting point for a more elaborate formulation.

The components of the Einstein tensor for the metric given by (\ref{a1}) 
with the scale factor in Eq.(\ref{a2}) are
\begin{eqnarray}
G_{00}&=& 3H^2\,=\,
\frac{\{a_2[6p_2(a_2t)^{\frac{7}{6}}+b_1(p_1+p_2a_2t)]\}^2}
{12(a_2t)^{\frac{7}{3}}(p_1+p_2a_2t)^2} \label{a6a} \\
G_{ij}&=&-(2\frac{\ddot{a}}{a}+H^2)g_{ij}
\,=\,\frac{36p_2^2(a_2t)^{\frac{7}{3}}+
3b_1^2(p_1+p_2a_2t)^2+2b_1(a_2t)^{\frac{1}{6}}
(-7p_1^2+4p_1p_2a_2t+11p_2^2a_2^2t^2)}{36t^2(a_2t)^{\frac{1}{3}}(p_1+p_2a_2t)^2}
g_{ij} \nonumber \\
\label{a6b}
\end{eqnarray}
Provided that we identify the source of the energy-momentum tensor as a 
collection of n real scalar fields, its general form is
\begin{eqnarray}
T_{\mu\nu}&=&\sum_{i=1}^n[\partial_\mu\phi_i\partial_\nu\phi_i+
g_{\mu\nu}[-\frac{1}{2}g^{\tau\rho}
\sum_{i=1}^n\partial_\tau\phi_i\partial_\rho\phi_i\,-\,
V(\phi_1,\phi_2,\cdot,\phi_n)] \nonumber \\
T_{00}&=&H\,=\,
\sum_{i=1}^n\frac{1}{2}\dot{\phi}_n^2\,+\,
V(\phi_1,\phi_2,\cdot,\phi_n) \label{a7a} \\
T_{ij}&=&
[\,\sum_{i=1}^n\frac{1}{2}\dot{\phi}_n^2\,-\,
V(\phi_1,\phi_2,\cdot,\phi_n)\,]g_{ij} 
\label{a7b} 
\end{eqnarray}
After using the Einstein equations,  we make the identification
\begin{eqnarray}
8\pi\,G
\sum_{i=1}^n\dot{\phi}_n^2&=&
G_{00}\,+\,\frac{G_{11}}{g_{11}}\,=\,
8\pi\,G(\,T_{00}+\frac{T_{11}}{g_{11}})\,=\,
\frac{36p_2^2 (a_2t)^{\frac{13}{6}}+7b_1(p_1+p_2a_2t)^2}
{18t^2(a_2t)^{\frac{1}{6}}(p_1+p_2a_2t)^2} \label{a8a} \\
16\pi\,G\,V&=&
G_{00}\,-\,\frac{G_{11}}{g_{11}}\,=\,
8\pi\,G(\,T_{00}-\frac{T_{11}}{g_{11}})\nonumber \\
&=&
\frac{72p_2^2(a_2t)^{\frac{7}{3}}+
3b_1^2(p_1+p_2a_2t)^2-b_1(a_2t)^{\frac{1}{6}}
(7p_1^2-22p_1p_2a_2t-29p_2^2a_2^2t^2)}
{18t^2(a_2t)^{\frac{1}{3}}(p_1+p_2a_2t)^2} \nonumber \\
&=&
\frac{4p_2^2a_2^2}{(p_1+p_2a_2t)^2}+
\frac{b_1^2a_2^2}{6(a_2t)^{\frac{7}{3}}}
-\frac{7b_1a_2^2}{18(a_2t)^{\frac{13}{6}}}
+\frac{2p_2b_1a_2^2}{(a_2t)^{\frac{7}{6}}(p_1+p_2a_2t)^2}
\label{a8b}
\end{eqnarray}
Eq.(\ref{a8a}) may be used to identify the scalars that act as the source 
of the Einstein equations
\begin{eqnarray}
&&\phi_1(t)\,=\,c_1(a_2t)^{-\frac{1}{12}} \label {a9a} \\
&&\phi_2(t)\,=\,c_2\,\ln{(p_1+p_2a_2t)} \label{a9b} \\
&&c_1\,=\,\sqrt{\frac{7b_1}{\pi\,G}}~,~~
c_2\,=\,\frac{1}{2\sqrt{\pi\,G}} \label{a9c}
\end{eqnarray}
Writing the potential $V$ in terms of these fields and satisfying the 
field equations
\begin{eqnarray}
\nabla_\mu\nabla^\mu\phi_1
\,-\,\frac{\partial\,V}{\partial\,\phi_1}
&=&-3\,H\,\dot{\phi_1}\,-\,\ddot{\phi_1} 
\,-\,\frac{\partial\,V}{\partial\,\phi_1}\,=\,0
\label{a10a} \\
\nabla_\mu\nabla^\mu\phi_2
\,-\,\frac{\partial\,V}{\partial\,\phi_2}
&=&-3\,H\,\dot{\phi_2}\,-\,\ddot{\phi_2} 
\,-\,\frac{\partial\,V}{\partial\,\phi_2} \label{a10b}\,=\,0 \label{a10b}
\end{eqnarray}
identifies $V$ as
\begin{eqnarray}
8\pi\,G\,V&=&
\,2p_2^2a_2^2\,\exp{(-2\frac{\phi_2}{c_2})}
\,+\,\frac{1}{12}b_1^2a_2^2(\frac{\phi_1}{c_1})^{28}
\,-\,\frac{7}{36}a_2^2b_1\,(\frac{\phi_1}{c_1})^{26} \nonumber \\
&+&p_2a_2^2b_1(\frac{\phi_1}{c_1})^{14}\,\exp{(-\frac{\phi_2}{c_2})}
 \label{a11a} 
\label{a11c} 
\end{eqnarray}
Then Eqs.(\ref{a10a},\ref{a10b}) are trivially satisfied for all values 
of the parameters, $a_2$, $b_1$, $p_1$, $p_2$. 

Next we 
will constrain these free parameters by phenomenological considerations 
and see if it gives a consistent and viable picture of the main lines of 
the cosmic history (except the baryonic matter and radiation). However 
before a phenomenological analysis it is necessary to identify which term 
in the above analysis corresponds to inflaton, which one to dark 
matter, and which one to dark energy. Before beginning the discussion it 
is worthwhile to note that both of $\phi_1$ and $\phi_2$ survive during 
all epochs of cosmic history. However only one of them is dominant in a 
given era of cosmic evolution. The inflaton term must be the one that is 
dominant and causes a huge cosmic acceleration at the time of inflation 
(i.e. at very small times). After examination of Eq.(\ref{a4}) and 
Eqs.(\ref{a8a},\ref{a8b}) we see that 
the dominant terms (of huge contributions) for early times are 
proportional to $(a_2t)^{-\frac{7}{3}}\propto\,\phi_1^{28}$. At 
intermediate times the 
dominant term is the term proportional to $-7p_1$ in (\ref{a4}) i.e. 
the $(a_2t)^{-\frac{13}{6}}\propto\,\phi_1^{26}$ term in 
(\ref{a8a},\ref{a8b}). 
Therefore $\phi_1$ accounts for both of the inflationary 
and dark matter dominated eras. 
At late times the dominant contribution is due to 
the terms of the form $\frac{1}{(p_1+p_2a_2t)^2}$ i.e. the terms 
containing $\phi_2$. Hence $\phi_2$ may be identified by 
dark energy. Because $\phi_1$ is identified by dark matter its 
coupling to standard model particles must be small. However it must 
have large enough coupling with standard model particles to generate 
enough 
reheating. This may be accomplished by assuming $\phi_1$ be 
electrically neutral and be color singlet. Even it may be taken to be 
a singlet under the whole $SU(3)_c\otimes\,SU(2)_L\otimes\,U(1)_L$ 
group of the standard model and couple to standard model particles 
indirectly (say through Higgs field) as in \cite{Shaposhnikov,hybrid}.
Another point would be a detailed study of the potential in (\ref{a11c}) 
especially to determine the effective range of $\phi_1$, in connection 
with its identification as dark matter, that is quite difficult due to the 
highly non-linear form of the potential. In fact all these points will 
arise when baryonic matter is included into the model and a more 
comprehensive and elaborate extension of this study is done in future.  
After these remarks we return to our main objective in 
the following paragraphs to check the phenomenological viability of the 
model. I put rough constraints on some of the free parameters, $a_2$, 
$b_1$, $p_1$, $p_2$ through a rough empirical analysis. The cosmological 
eras that I employ to put constraints are the inflationary era, the 
present day, the onset of matter dominated era, and the time of 
reionization. I also consider the time of matter - radiation decoupling 
time.

\section{compatibility with observations}
The value of the present value of scale factor is taken to be one 
by convention. This implies
\begin{eqnarray}
&&1=a_0=a(t_0)\,=\,
(p_1+p_2a_2t_0)
\exp{[-b_1(a_2t_0)^{-\frac{1}{6}}]}\label{d1} \\
&&\Rightarrow~~~~~~\
(p_1+p_2a_2t_0) \,=\,
\exp{[b_1(a_2t_0)^{-\frac{1}{6}})]}\,=\, \beta\,>\,1
\label{a20}
\end{eqnarray}
where $\beta$ is some constant to be determined from observational data. 
We exclude the case $\beta=1$ since it corresponds to infinite 
time for the present age of the universe. 
Next consider the observational values of the present value of the Hubble 
constant $H(t=t_0)=H_0$ and the age of the universe $t_0$. 
The observational values of 
$H_0=\frac{h}{(9.777752\,Gyr)}\simeq\,\frac{1}{13.5802\,Gyr}$, and 
$t_0=13.69\pm0.13\,Gyr$ given by Particle Data Group (PDG) \cite{PDG} 
gives 
\begin{equation}
0.998\,<\,H_0t_0\,<\,1.018 \label{a12aa}
\end{equation}
This implies $H_0t_0\simeq\,1$. Although the $H_0$ and $t_0$ values in 
(\ref{a12aa}) are the most standard 
values, there are different observational values for $H_0$ and $t_0$ as 
well. For example Reese et. al. finds a value of $H_0$ smaller than the 
PDG value by approximately $16\%$ \cite{Reese} although it may be 
ascribed to underestimation of the SZE/X-ray derived distances. The 
central values of the age of 
the universe derived from other methods as well differ from PDG value. For 
example $t_0$ derived by 
the age determinations of elements by radioactive decay ratio method give 
the age of Milky Way ranging from 12.3 to 17.3 Gyr \cite{element-age},
the radioactive dating of old stars give values in the range 11 to 20.2 
Gyrs  \cite{old-star-age}, the age of the oldest star cluster ranges from 
8.5 to 16.3 Gyrs \cite{oldest-star-cluster-age}.
Another point to mention is that the PDG value of $t_0$ is determined from 
$\Lambda$CDM model.
Therefore it is better to be more open minded to be about the value of 
$H_0t_0$,  and hence in the following I take 
\begin{eqnarray}
H_0t_0&=&
\frac{p_2a_2t_0}{p_1+p_2a_2t_0}\,+\,
\frac{1}{6}b_1a_2^{-\frac{1}{6}}t_0^{-\frac{1}{6}}
\,=\,\xi\,\sim\,1 \nonumber \\
&&\Rightarrow~~~~
p_2a_2t_0\,=\,\beta(\xi-\frac{1}{6}\ln{\beta})~,~~~
p_1\,=\,
\beta(1-\xi+\frac{1}{6}\ln{\beta})
\label{c1}
\end{eqnarray}
where Eqs. (\ref{a3a}) and (\ref{a20}) are employed. 

One may obtain a  constraint on the value of $\beta$ by using 
the cosmic deceleration 
period (in the matter dominated era).
Eq.(\ref{a4}) suggests that at the matter dominated era
\begin{eqnarray}
&&-7p_1(a_2t_m)^{\frac{1}{6}}+p_1b_1\,<\,0
~~~~\Rightarrow~~~~
t_m\,>\,\frac{1}{a_2}\left(\frac{b_1}{7}\right)^6 \label{b1a} \\
&&-7p_1\,+\,5p_2a_2t_m\,<\,0
~~~~\Rightarrow~~~~t_m\,<\,
\frac{7p_1}{5p_2a_2} \label{b1b}\\
&&\Rightarrow~~~~
\frac{1}{a_2}\left(\frac{b_1}{7}\right)^6 \,<\,t_m\,<\,
\frac{7p_1}{5p_2a_2} \nonumber \\
&& \left(\frac{1}{7}\ln{\beta}\right)^6\,<\,\gamma_m\,<\,
\frac{7(1-\xi+\frac{1}{7}\ln{\beta})}{5(\xi-\frac{1}{6}\ln{\beta})}
\label{b1c}
\end{eqnarray}
where $t_m$ denotes the time of deceleration in the matter dominated 
era, and $\gamma_m=\frac{t_m}{t_0}$. Note that the inequalities above do 
not 
saturate i.e. the lower and the upper values in the inequalities are not 
infinitesimally close to the initial and final times of cosmic 
deceleration. In 
fact a more stringent bound on $\beta$ and the time of the onset of 
cosmic acceleration in the dark energy dominated era may be obtained . It 
is evident from 
(\ref{a4}) that $t^\prime\,=\,\frac{7p_1}{5p_2a_2}$ is greater than the 
time of onset of dark energy dominated era, $t_d$ i.e. 
$t_d\,=\,\alpha\,t^\prime$, 
$\alpha\,<\,1$ because of the additional terms contributing to the 
denominator of Eq.(\ref{a4}) in addition to those considered in 
Eqs.(\ref{b1a},\ref{b1b}). Hence at the onset of cosmic acceleration one 
may write
\begin{eqnarray}
&&b_1(p_1+p_2a_2t_d)+(a_2t_d)^{\frac{1}{6}}(-7p_1+5p_2a_2t_d)\,=\,0
~~~~~\Rightarrow 
\nonumber \\
&&b_1(p_1+p_2a_2\alpha\,t^\prime)
+(a_2\alpha\,t^\prime)^{\frac{1}{6}}(-7p_1+5p_2a_2\alpha\,t^\prime)
\nonumber \\
&=&\frac{b_1}{7}(7p_1-5p_2a_2t^\prime)\,+\,b_1[\frac{5}{7}p_2a_2t^\prime\,+\,
p_2a_2t^\prime]\,+\,
(a_2\alpha\,t^\prime)^{\frac{1}{6}}
[-7p_1
\,+\,5p_2a_2t^\prime\,+\,5p_2a_2(\alpha-1)t^\prime]\nonumber \\
&=&b_1(\frac{5}{7}\,+\,\alpha)p_2a_2t^\prime
\,+\,5p_2a_2(\alpha-1)t^\prime(a_2\alpha\,t^\prime)^{\frac{1}{6}}
\,=\,0
\nonumber \\
&&\Rightarrow~~~~~b_1(\frac{5}{7}+\alpha)
+5(\alpha-1)(a_2\alpha\,t^\prime)^{\frac{1}{6}}\,=\,0 \nonumber \\
&&\Rightarrow~~~~~\frac{b_1}{a_2^{\frac{1}{6}}}(\frac{5}{7}+\alpha)
+5(\alpha-1)t_d^{\frac{1}{6}}\,=\,0~~~~~~\Rightarrow~~~~
\ln{\beta}\,=\,\frac{
5\left(1-\alpha\right)}{\frac{5}{7}+\alpha}\gamma_d^{\frac{1}{6}}
\label{b5}
\end{eqnarray}
where $\gamma_d=\frac{t_d}{t_0}$, and $t^\prime$ is the time satisfying 
$-7p_1+5p_2a_2t^\prime=0$.
We know that $t_d\,<\,t_0$. The observational data analyzed in the 
context of 
$\Lambda$CDM model and dynamical dark energy models with a moderate 
dependence on redshift
gives 
$\frac{t_d}{t_0}\simeq\,\frac{1}{2}$ 
\cite{onset1}. The fact that there is 
no significant disagreement of the  
$\Lambda$CDM model with data implies that the value of $t_d$ should not be 
too different from this value. If one takes 
$\frac{t_d}{t_0}\,=\,\frac{1}{2}$ 
$\left(\frac{t_d}{t_0}\right)^{\frac{1}{6}}\,\simeq\,0.89$ while for
$\frac{t_d}{t_0}\,=\,\frac{1}{100}$ 
$\left(\frac{t_d}{t_0}\right)^{\frac{1}{6}}\,\simeq\,0.464$. Therefore it 
is safe to say that 
$\gamma_d=\left(\frac{t_d}{t_0}\right)^{\frac{1}{6}}\,\sim\,1$ for 
reasonable 
values of $\gamma_d$. Then one may get an idea of the magnitude of 
$\ln{\beta}$ for a few values of $\alpha$ by using Eq.(\ref{b5})
\begin{eqnarray}
&&\alpha\,=\,1~~~~\Rightarrow~~~~
ln{\beta}\,=\,0
~~~~~~\Rightarrow~~~~~\beta\,=\,1 
\label{c5a} \\
&&\alpha\,=\,\frac{9}{10}~~~~\Rightarrow~~~~
ln{\beta}\,=\,
\frac{35}{113}\gamma_d^{\frac{1}{6}}\,\sim\,
\frac{35}{113}
~~~~~~\Rightarrow~~~~~\beta\,\sim\,1.63 
\label{c5b} \\
&&
\alpha\,=\,\frac{5}{10}~~~~\Rightarrow~~~~
ln{\beta}\,=\,\frac{2.5}{\frac{17}{4}}
\gamma_d^{\frac{1}{6}}\,\sim\,
\frac{2.5}{\frac{17}{4}}
~~~~~~\Rightarrow~~~~~\beta\,\sim\,7.8
\label{c5c} \\ 
&&
\alpha\,=\,\frac{1}{10}~~~~\Rightarrow~~~~
ln{\beta}\,=\,\frac{31.5}{5.7}\gamma_d^{\frac{1}{6}}\,\sim\,
\frac{31.5}{5.7}
~~~~~~\Rightarrow~~~~~\beta\,\sim\,251 
\label{c5d} \\
&&\alpha\,=\,0~~~~\Rightarrow~~~~
ln{\beta}\,=\,
\frac{5}{\frac{5}{7}}\gamma_d^{\frac{1}{6}}\,\sim\,7
~~~~~~\Rightarrow~~~~~\beta\,\sim\,1097
\label{c5e} 
\end{eqnarray}
It is evident that in any case
\begin{equation}
\ln{\beta}\,<\,7 \label{f1}
\end{equation}
In the following paragraphs we take this as an upper bound on the 
values of $\ln{\beta}$ and we do not consider higher values unless it 
seems necessary for the sake of completeness.

Now we derive a lower bound on the value of $\beta$ by using the 
$G_{00}/\frac{\ddot{a}}{a}$ at present time. Note that we use 
$G_{00}/\frac{\ddot{a}}{a}$ rather than the equation of state for 
dark energy since the dark energy and dark matter fields are mixed in the 
energy-momentum tensor so that it becomes impossible to entangle the dark 
energy and dark matter contributions properly in this case.   
\begin{eqnarray} 
\left(\frac{G_{00}}{\frac{\ddot{a}}{a}}\right)_{t=t_0}&=&
-\left(\frac{8\pi\,G\rho}{\frac{1}{6}8\pi\,G(\rho+3p)}\right)_{t=t_0}\,=\,
\frac{3[6p_2(a_2t_0)^{\frac{7}{6}}+b_1(p_1+p_2a_2t_0)]^2}
{b_1(p_1+p_2a_2t_0)[b_1(p_1+p_2a_2t_0)
+(a_2t_0)^{\frac{1}{6}}(-7p_1+5p_2a_2t_0)]} \nonumber \\
&=&-\left(\frac{3}{ln{\beta}}\right)\frac{36\xi^2}
{(\ln{\beta}+7-12\xi)}
\label{c2}
\end{eqnarray}
The PDG values 
$-1.14\,<\,\omega_{dark-e}\,<\,-0.95$, 
$0.21\,<\,\Omega_m=\frac{\rho_c}{\rho_m}\,<\,0.26$,
$\Omega_{dark-e}=\frac{\rho_c}{\rho_{dark}}\,\simeq\,0.74$ may be used to 
calculate (\ref{c2}).
The corresponding observational value of the ratio is 
\begin{equation}
3.9\,<\,\left(\frac{G_{00}}{\frac{\ddot{a}}{a}}\right)_{t=t_0}\,=\,
\frac{6}{1+3\Omega_{dark-e}\omega_{dark-e}}\,<\,5.5
\label{c3a}
\end{equation}
However there are studies with a wider range for current equation of state 
and density parameter from the analysis of SNe data 
alone \cite{wider-omega}
\begin{eqnarray}
&&-1.7\,<\,\omega_{dark-e}\,<\,-0.50~,~ 
0.23\,<\,\Omega_m=\frac{\rho_c}{\rho_m}\,<\,0.37 \label{c3aa} \\
&&
-6/(1+3(-1.7)0.77)=2.05\,<\,
\left(\frac{G_{00}}{\frac{\ddot{a}}{a}}\right)_{t=t_0}=
-\frac{6}{1+3\Omega_\Lambda\omega_\lambda}\,<\,\infty
\label{c3b}
\end{eqnarray}
Although the infinity is unphysical I do not know a stringent and definite 
upper bound to be replaced by the $\infty$ in (\ref{c3b}). Therefore I 
keep it as infinity.  However one may replace $\infty$ by a large 
enough value. For example \cite{Komatsu2009} gives 
upper bound $\beta\simeq23$. 
We plot 
$\left(\frac{G_{00}}{\frac{\ddot{a}}{a}}\right)_{t=t_0}$ 
versus $\beta$ for various values of $\gamma$ and $\xi\sim\,1$. The 
results are given in Table \ref{table-G00-ac-present}. 
\begin{table}[h] 
\begin{tabular}{|c|c|c|} \hline
$\xi$&
$\left(\frac{G_{00}}{\frac{\ddot{a}}{a}}\right)_{t=t_0}$&$\beta$ \\
\hline 0.5&3.9~ --- ~5.5&\mbox{none}
\\
\hline 
0.5&2.05~ --- ~23
&
2.6~---~3
\\
\hline 
0.8&3.9~ --- ~5.5&\mbox{none}\\
\hline 
0.8&2.05~ --- ~23&2.14~---~12.2\\
\hline 
0.9&3.9~ --- ~5.5&\mbox{none}\\
\hline 
0.9&2.05~ --- ~23
&1.89~---~43.68\\
\hline 
1&3.9~ --- ~5.5&3.95~---~144.4\\
\hline 
1&2.05~ --- ~23
&1.804~---~147.4\\
\hline 
1.2&3.9~ --- ~5.5&3.55~---~1632.16\\
\hline 
1.2&2.05~ --- ~23
&1.757~---~1635.05\\
\hline 
1.5&3.9~ --- ~5.5&3.56~---~4.34\\ 
\hline
1.5&2.05~ --- ~23
&1.77~---~6.8 \\
\hline 
\end{tabular}
\caption[b]{The allowed values of $\beta$ for two intervals of
$\left(G_{00}/\frac{\ddot{a}}{a}\right)_{t=t_0}$,and various values of 
$\xi$}  
\label{table-G00-ac-present}
\end{table} 
One sees that the 
values of $\beta$ compatible with (\ref{c3a}) are greater than 2.2
while the lower bound on $\beta$ for (\ref{c3b}) with $\beta_u=9$ are 
greater than 2.6. 

\begin{figure} \includegraphics[scale=0.9]{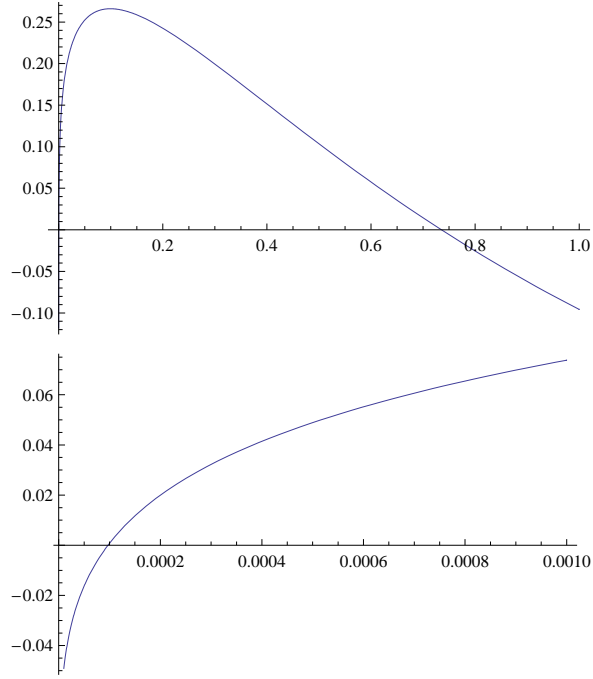} 
\caption[]{ $\left(\frac{G_{11}}{g_{11}}/G_{00}\right)$ 
versus
$\gamma=\frac{t}{t_0}$ graph for $\beta=50$, $\xi=1$}
\label{fig:w}
\end{figure}

As a complimentary analysis one may determine the ratio $G_{ij}/G_{00}$. 
Consider $\frac{G_{11}}{g_{11}}/G_{00}$ at present time
\begin{eqnarray} 
\left(\frac{\frac{G_{11}}{g_{11}}}{G_{00}}
\right)_{t=t_0}&=&
-\frac{12p_2^2(a_2t_0)^{\frac{7}{3}}+b_1^2(p_1+p_2a_2t_0)^2+\frac{2}{3}b_1
(a_2t_0)^{\frac{1}{6}}(-7p_1^2+4p_1p_2a_2t_0+11a_2^2p_2^2t_0^2)}
{[6p_2(a_2t_0)^{\frac{7}{6}}+b_1(p_1+p_2a_2t_0)]^2}
\nonumber \\
&=&-\frac{18\xi^2-(7-12\xi)\ln{\beta}+(\ln{\beta})^2}
{54\xi^2}
\label{c7}
\end{eqnarray}
A $\left(\frac{G_{11}}{g_{11}}/G_{00}\right)_{t=t_0}$ versus $\beta$ 
graph may be plotted for various values of $\xi$. I take 
$-0.315\,<\,\Omega_{dark-e}\omega_{dark-e}\,<\,-1.3629$ 
by using Eq.(\ref{c3aa}). The corresponding allowed 
range of values of 
$\beta$ for some values of $\xi$ are given below 
\begin{eqnarray}
&&\xi\,=\,0.8~~~~~~\beta\,=\,0.8~---~16.8
\nonumber \\
&&\xi\,=\,0.9~~~~~~\beta\,=\,0.82~---~54
\nonumber \\
&&\xi\,=\,1~~~~~~\beta\,=\,0.8~---~180 \nonumber \\
&&\xi\,=\,1.2~~~~~~\beta\,=\,0.83~---~2\,\times\,10^3 \label{c8}
\end{eqnarray}
In fact we should exclude the values of $\beta$ smaller than one 
given above
because of the definition of $\beta$ in (\ref{a20}).
The values of $\beta$ above are 
barely consistent with 
the more stringent bounds in Table \ref{table-G00-ac-present} for 
$\xi=0.8,\,0.9$  and are 
consistent in the upper range for the others. The 
$\left(\frac{G_{11}}{g_{11}}/G_{00}\right)$ value corresponds to the 
effective equation of state of the dark fluid consisting of dark energy 
and dark matter. Therefore it is useful to give its general time 
dependence as well. 
\begin{eqnarray}
\frac{\frac{G_{11}}{g_{11}}}{G_{00}}
&=&
-\frac{12p_2^2(a_2t)^{\frac{7}{3}}+b_1^2(p_1+p_2a_2t)^2+\frac{2}{3}b_1
(a_2t)^{\frac{1}{6}}(-7p_1^2+4p_1p_2a_2t+11a_2^2p_2^2t^2)}
{[6p_2(a_2t)^{\frac{7}{6}}+b_1(p_1+p_2a_2t)]^2}
\nonumber \\
&=&-\frac{\frac{\ln{\beta}}{12\gamma^{\frac{7}{3}}}
+\frac{
(\xi-\frac{1}{6}\ln{\beta})^2}
{[1-(1-\gamma)\xi+
\frac{(1-\gamma)}{6}
\ln{\beta}]^2}+\frac{11(
(\xi-\frac{1}{6}\ln{\beta})^2\ln{\beta}}{18\gamma^{\frac{7}{6}}
(1-(1-\xi)\gamma+\frac{(1-\gamma)}{6}\ln{\beta})}-
\frac{
7(1-\xi+\frac{1}{6}\ln{\beta})\ln{\beta}}{18\gamma^{\frac{13}{6}}
(1-(1-\xi)\gamma+\frac{(1-\gamma)}{6}\ln{\beta})}}
{3\left(\gamma^{-\frac{7}{6}}\frac{\ln{\beta}}{6}
+\frac{\xi-\frac{1}{6}\ln{\beta}}
{(1-(1-\xi)\gamma+\frac{(1-\gamma)}{6}\ln{\beta})}\right)^2}
\label{af7}
\end{eqnarray}
where $\gamma=\frac{t}{t_0}$. As we shall remark later in this section a 
general analysis of this 
effective equation state at an arbitrary redshift is quite difficult due 
to the highly nonlinear form of the above equation. However one get an 
idea of its general variation by the inspection of Figure \ref{fig:w} for 
$\beta=50$, $\xi=1$. 

\begin{figure}
\includegraphics[scale=0.9]{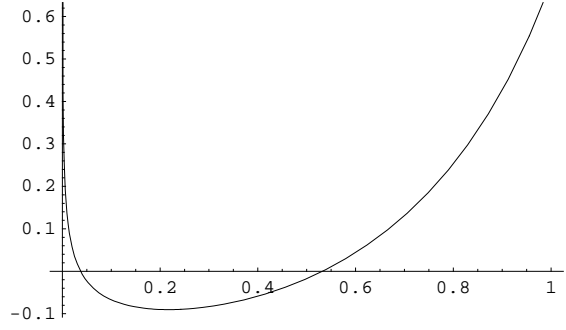}
\caption[]{
$\frac{\ddot{a}}{a}/G_{00}$ versus
$\gamma=\frac{t}{t_0}$ graph for $\beta=50$, $\xi=1$}
\label{fig:acc}
\end{figure}

The general form of the cosmic history must have 
a cosmic acceleration era corresponding to the time of inflation that 
is followed by an era of deceleration at the matter dominated era, and 
finally by the present time acceleration era. Moreover the redshift values 
and ages for these eras must 
coincide with the 
observational data \cite{PDG} at least at the order of magnitude to have 
at 
least an approximately realistic model. For this purpose we draw 
$\frac{\ddot{a}}{a}/G_{00}$ versus time for $\xi$=0.8,, 1, 1.2; 
$\beta$=2, 5, 10, 20, 50, 100, 200, 500, 1000, 2000, 10000, 3000, 40000 by 
using
\begin{eqnarray} 
\frac{\frac{\ddot{a}}{a}}{G_{00}}
&=&
-\frac{(a_2)^{\frac{1}{6}}b_1(p_1+p_2a_2t)[\frac{b_1}{(a_2t)^{\frac{1}{6}}}
(p_1+p_2a_2t)+(-7p_1+5p_2a_2t)]} {3(a_2t)^{\frac{1}{3}}[6p_2a_2t
+\frac{b_1}{(a_2t)^{\frac{1}{6}}}(p_1+p_2a_2t)]}
\nonumber \\
&=&-\left(\frac{ln{\beta}}{3\gamma^{\frac{1}{6}}}\right)
\frac
{A[(\gamma^{-\frac{1}{6}}\ln{\beta})A
-7(1-\xi+\frac{1}{6}\ln{\beta})+
5\gamma\xi-\frac{5}{6}\gamma\ln{\beta}]}
{6(\xi-\frac{1}{6}\ln{\beta})\gamma+\gamma^{-\frac{1}{6}}
\ln{\beta}[1-(1-\gamma)\xi+\frac{1-\gamma}{6}\ln{\beta}]}
\label{f2} \\
&&A\,=\,[1-(1-\gamma)\xi+\frac{(1-\gamma)}{6}\ln{\beta}] \nonumber 
\end{eqnarray}
where $\gamma=\frac{t}{t_0}$. 
One may get a sense of  the general form of the evolution in Figure 
\ref{fig:acc} for $\beta=50$, $\xi=1$. 
I do not give the other plots not to make the paper too crowded.
The all values of the times of  the start and 
the end of cosmic 
deceleration that are not wholly excluded by data are given in 
Table \ref{table-tmi-tmf}
below. 
It seems that this analysis prefers lower values of 
$\beta$. Note 
that we should not expect a good match between the values 
obtained here and the observational data for the time of the start of the 
deceleration period because at that time the radiation has a major 
contribution and we neglect the contribution of radiation in this study.
\begin{table}[h] 
\begin{tabular}{|c|c|c|c||c|c|c|c|} \hline
$\xi$&$\beta$&$\gamma_{mi}$&$\gamma_{mf}$&
$\xi$&$\beta$&$\gamma_{mi}$&$\gamma_{mf}$\\
\hline 
0.8&2&$10^{-6}$&
0.5&0.8& 5 &$1.5\times10^{-4}$&0.7\\
\hline 
0.8&10&
$1.2\times10^{-3}$&0.9&
0.8&200&
0.165&0.947\\
\hline 
0.8&500&0.28&0.89&
0.8&1000&
0.35&0.867\\
\hline 
0.8&2000&0.58&0.91&-&-&-&-\\
\hline 
1&2&
$9.5\times10^{-7}$&0.13&
1&5&
$1.4\times10^{-4}$&0.26\\
\hline 
1&10 & 
$1.3\times10^{-3}$&0.34&
1&20 & 
$6.6\times10^{-3}$&0.415
\\
\hline 
1&50 &0.04&0.45&
1&100&0.116&0.7 \\
\hline 
1&1000 &0.48&0.945&
1&2000&0.52&0.916 \\
\hline 
\end{tabular}
\caption[b]{The times of start, $\frac{t_{mi}}{t_0}$ and end, 
$\frac{t_{mf}}{t_0}$ of the cosmic deceleration  for different 
values of $\beta$}  \label{table-tmi-tmf}
\end{table}

Now we compare the data and the predictions of this model for different 
redshifts and times, and hope, 
at least, an order of magnitude agreement. First consider the time of the 
starting of cosmic acceleration. 
This is the time where the deceleration changes into acceleration, 
hence the acceleration of the cosmic expansion is zero i.e. the 
numerator of (\ref{a4}) is zero, namely 
\begin{eqnarray}
&&f\,=\,\frac{1-\gamma_d}{6}x^2+[1-(1-\gamma_d)\xi-\frac{7}{6}\gamma_d^{\frac{1}{6}}
-\frac{5}{6}\gamma_d^{\frac{7}{6}}]x-7(1-\xi)\gamma_d^{\frac{1}{6}}\,=\,0
\label{c10} \\
&&~~\gamma_d\,=\,\frac{t_d}{t_0}~~~,~~~x\,=\,\ln{\beta} \nonumber \\
\end{eqnarray}
Because of its highly non-linear form this equation I could not 
analytically solve this equation. However after plotting $f$ versus x for 
various values of $\xi$ and $\gamma_d$ and determining the location of 
zeros one may get some information. The 
result is given in Table \ref{table-transition-acc-dec}. 
\begin{table}[h] 
\begin{tabular}{|c|c|c||c|c|c|} \hline
$\xi$&$\gamma_d$&$x=\ln{\beta}$&
$\xi$&$\gamma_d$&$x=\ln{\beta}$
\\
\hline 
0.8&0.85&2 or 26&
0.8&0.8&2 or 25\\
\hline 
0.8&0.75&2 or 20&
0.8&0.7&2 or 16\\
\hline 
0.8&0.5&0.7 or 9&
0.8&0.4&0.2 or 7.8\\
\hline 
0.8&0.1&1 or 5&
0.8&$10^{-10}$&-1.2 or 0.2\\
\hline 
0.8&$10^{-20}$&-1.2 or 0&
1&0.85&5 or 35\\
\hline 
1&0.8&5 or 24&
1&0.75&5 or 17\\
\hline 
1&0.7&5 or 15&
1&0.5&3.6 or 7.2\\
\hline 
1&0.3&1.9 or 5.5&
1&0.1&0.5 or 4.5\\
\hline 
1&$10^{-10}$&0.06 or 0.09&
1&$10^{-20}$&0 or 0.0033\\
\hline 
1.2&0.85&7 or 33&
1.2&0.8&7 or 22\\
\hline 
1.2&0.75&8.7 or 15.7&
1.2&0.5&none\\
\hline 
\end{tabular}
\caption[b]{The allowed values of $ln{\beta}$ for various values 
of $\xi$ and $\gamma_d$}  \label{table-transition-acc-dec}
\end{table}
Keeping these values in mind now we may find the redshift values and 
the time of onset of current cosmic acceleration, $t_d$ predicted by this 
model and compare the observational values given in literature.
Consider 
$a_0/a(t_d)$
\begin{equation}
\frac{a_0}{a(t_d)}
\,=\,\frac{\exp{[b_1(a_2t_d)^{-\frac{1}{6}}]}}{(p_1+p_2a_2t_d)}
\,=\,\frac{\beta^{-1+\gamma_d^{-\frac{1}{6}}}}
{
(1-\xi)
+\frac{1}{6}\ln{\beta}+(\xi-\frac{1}{6}\ln{\beta})\gamma_d}
\label{a22}
\end{equation}
The analysis of cosmic data \cite{onset} 
gives the redshift and 
time of onset of dark energy dominated era, respectively, in the ranges 
$z\,=\,0.66\,-\, 1.21$. 
$t_d\,=\,(5.7\,-\,8.5)\,Gyr$. 
The allowed intervals of $\beta$ in (\ref{a22}) where $\frac{a_0}{a(t_d)}$ 
is in 
the range $\,1.66\,--\, 2.21$ 
for the phenomenologically 
relevant values of $\xi$ and $\gamma_d=\frac{t_d}{t_0}$ for $\xi$= 
0.8, 1, 1.2 and $\gamma_d$=0.1, 0.4, 0.5, 0.6, 0.7, 0.8 
may be found in Table \ref{table-atd-b}. 
\begin{table}[h] 
\begin{tabular}{|c|c|c|} \hline
$\xi$&$\gamma_d$&$\beta$ \\
\hline 
0.8&0.1&none\\
\hline 
0.8&0.4&
0.15~---~123
\\
\hline 
0.8&0.5&1~---~$7\times10^3$
\\
\hline 
0.8&0.6&$2\times10^3~---~2.1\times10^6$ and 
$1.8\times10^{-3}~---~2.5\times10^{-2}$ \\
\hline 
0.8&0.8&
$5\times\,10^{12}~---~2.5\times\,10^{18}$ and 
$7\times\,10^{-9}~---~3\times\,10^{-7}$ \\
\hline 
1&0.1&none\\
\hline 
1&0.4&none
\\
\hline 
1&0.5&4~---~1400\\
\hline 
1&0.6&
$130~---~7\times10^{5}$ and 
$1.2\times\,10^{-2}~---~1$\\
\hline 
1&0.7&
$7.5\times10^5~---~10^{10}$ and 
$1.7\times\,10^{-4}~---~5\times\,10^{-3}$\\
\hline 
1&0.8&
$1.5\times\,10^{12}~---~\,10^{18}~~\mbox{and}~~~
5\times\,10^{-8}~---~2.8\times\,10^{-6}$\\
\hline 
1.2&0.1&none\\
\hline 
1.2&0.4&none\\
\hline 
1.2&0.5&barely 30
\\
\hline 
1.2&0.6&
$
0.5~---~2\times\,10^{5}$ \\
\hline
1.2&0.7&
$9\times10^{4}~---~3.5\times\,10^{9}$ and
$10^{-3}~---~9\times\,10^{-2}$\\
\hline
1.2&0.8&
$10^{12}~---~5\times\,10^{17}$ and
$2.6\times\,10^{-7}~---~2.1\times\,10^{-5}$\\
\hline 
\end{tabular}
\caption[b]{The allowed range of values of $\beta$ for various values of
$\xi$, $\gamma$ with $\frac{a_0}{a(t_d)}$ in the range $\,1.66\,--\, 2.21$}  
\label{table-atd-b}
\end{table}
Table \table-atd-b} tells us that $\gamma_d=0.1$ is inconsistent with 
data. 
Comparison with Eq.(\ref{c8}) and Table \ref{table-G00-ac-present} implies 
that the $\gamma_d$ values in the 
range 0.4 - 0.6 are consistent with redshift data \cite{onset} and the 
time of the onset of the cosmic acceleration for the phenomenologically 
relevant values of $\xi$ in the range 1.2 - 0.8. 

An important point is to 
be mentioned at this point:
Note that the values $z\,=\,0.66\,-\, 1.21$,  
$t_d\,=\,(5.7\,-\,8.5)\,Gyr$ in \cite{onset1} are derived by the 
assumption 
that the Hubble constant at scale factor $a(t)$ may be expressed as
\begin{equation} 
H(a)\,=\,H_0\,[
\frac{\Omega_m}{a^3}+
\frac{\Omega_r}{a^4}+
\frac{\Omega_k}{a^2}+
\frac{\Omega_\Lambda}{a^{3(1+\omega_{eff}(a))}}\,]^{\frac{1}{2}}
\label{f4}
\end{equation}
In principle one may define an effective equation of state 
$\omega_{eff}$ as in \cite{Sola2} when dark matter and dark energy are 
coupled. 
However in this model the contributions of dark matter and dark energy are 
not only coupled they are mixed. Therefore their contributions
can not be separated from each other properly. Moreover dark matter in 
this 
model is not dust-like (it only mimics a dust in the matter dominated era) 
while the matter in the above equation is dust-like. Furthermore 
in \cite{onset1} and similar studies an equation of state for dark energy 
of the form 
$\omega(a)\,=\,-\omega_0+\omega_1(1-a)$ 
or similar forms are employed. Let 
alone that a proper equation of state for dark energy in this model can 
not be defined a common equation of state for dark energy and dark 
matter is highly nonlinear as seen before in Eq.(\ref{af7}) 
\begin{eqnarray}
\frac{\frac{G_{11}}{g_{11}}}{G_{00}}
&=&-\frac{\frac{\ln{\beta}}{12\gamma^{\frac{7}{3}}}
+\frac{
(\xi-\frac{1}{6}\ln{\beta})^2}
{[1-(1-\gamma)\xi+
\frac{(1-\gamma)}{6}
\ln{\beta}]^2}+\frac{11(
(\xi-\frac{1}{6}\ln{\beta})^2\ln{\beta}}{18\gamma^{\frac{7}{6}}
(1-(1-\xi)\gamma+\frac{(1-\gamma)}{6}\ln{\beta})}-
\frac{
7(1-\xi+\frac{1}{6}\ln{\beta})\ln{\beta}}{18\gamma^{\frac{13}{6}}
(1-(1-\xi)\gamma+\frac{(1-\gamma)}{6}\ln{\beta})}}
{3\left(\gamma^{-\frac{7}{6}}\frac{\ln{\beta}}{6}
+\frac{\xi-\frac{1}{6}\ln{\beta}}
{(1-(1-\xi)\gamma+\frac{(1-\gamma)}{6}\ln{\beta})}\right)^2}
\label{f7}
\end{eqnarray}
An inspection of $[a(t)]^{-1}$ versus $\gamma_0=\frac{t}{t_0}$ graphs show 
that in the low redshift range z= 05 - 2 one may approximately take 
$[a(t)]^{-1}$ proportional to $\gamma$. In other words one may get the 
form of $\omega(a)$ for low redshifts by simply replacing $\gamma$ in 
(\ref{f7}) by 1+z. It is evident that this relation is quite nonlinear in 
z. For higher redshift values the relation between $\gamma$ and 
(z+1)$^{-1}$ also becomes non-linear making the form of $\omega(a)$ even 
more complicated. Therefore in order to see the degree of the 
compatibility of this model with data in a more precise way it is 
necessary to repeat the analysis of data in \cite{onset,onset1} with 
keeping these points in mind. Only then one can say some definite 
conclusion on the degree of the agreement between this model and 
observational data. In any case I think the rough analysis given in this 
paper is enough to consider this model as viable toy model in the 
direction of unification of all eras of cosmic history. In fact all I have 
mentioned in the context of the analysis of the onset of cosmic 
acceleration data is true for the analysis of data for equation of state 
of 
dark energy \cite{Amanullah}, density parameters \cite{wider-omega}, and 
the analysis of 
data on time reionization and time of matter-radiation decoupling 
\cite{PDG,Komatsu} discussed below.

Before continuing the comparison of the model with observational 
analysis for the times of reionization and decoupling now I want to 
consider the inflationary era because there is no baryonic matter or 
radiation effect in this era, this toy model is 
expected to be most similar to the reality in this era in the 
context of this model. 
It is evident from (\ref{a4}) that, at 
the time of inflation, \begin{equation}
b_1(p_1+p_2a_2t)+(a_2t)^{\frac{1}{6}}(-7p_1+5p_2a_2t)\,>\,0 \label{a12}
\end{equation}
This condition is satisfied for very large and very small $a_2t$'s. We 
identify the very small $a_2t$ values that satisfy (\ref{a12}) as 
inflationary 
times, and at small times (\ref{a12}) is guaranteed if we take
$a_2t_i\,\ll\,1 \label{a13}$ where the subindex $i$ refers to inflation. 
In fact a more stringent bound may be obtained from the 
slow-roll parameter $\frac{\dot{H}}{H^2}$ in (\ref{a3b})
\begin{equation}
\frac{\dot{H}(t_i)}{H^2(t_i)} 
\,\simeq\,-\frac{7\gamma_i^{\frac{1}{6}}}{\ln{\beta}}
\,\ll\,1 
~~~\Rightarrow~~~~\gamma_i=\frac{t_i}{t_0}\,\ll\,1 \label{a13}
\end{equation}
where we have used the fact that the $p_1$ term 
in (\ref{a3b}) is the leading term in the inflationary period.
Then 
\begin{eqnarray}
\frac{a(t_{is})}{a(t_{il})}&=&
\frac{(p_1+p_2a_2t_{is})\exp{\{-b_1[
(a_2t_{is})^{-\frac{1}{6}}-
(a_2t_{il})^{-\frac{1}{6}}]\}}}
{(p_1+p_2a_2t_{il}} \nonumber \\
&=&\left(\frac{1-(1-\gamma_{is})\xi+\frac{(1-\gamma_{is})}{6}\ln{\beta}}
{1-(1-\gamma_{il})\xi+\frac{(1-\gamma_{il})}{6}\ln{\beta}}\right)
\exp{\{-\frac{b_1}{a_2^{\frac{1}{6}}}[
t_{is}^{-\frac{1}{6}}-
t_{il}^{-\frac{1}{6}}]\}}
\,\simeq\,
\exp{\{-\frac{b_1}{a_2^{\frac{1}{6}}}[
t_{is}^{-\frac{1}{6}}-
t_{il}^{-\frac{1}{6}}]\}} \nonumber \\
\label{a16a}
\end{eqnarray}
where $t_{is}$ and $t_{il}$ are the times of the start and end of 
inflation, respectively.
If we 
assume 60 e-fold expansion and $t_{is}= \,=\,10^{-36}\,sec$, 
$t_{il}\,=\,10^{-32}\,sec$
\begin{eqnarray}
\frac{a(t_{is})}{a(t_{il})}&\simeq&
\exp{[-b_1\,
(a_2)^{-\frac{1}{6}}
10^6(sec)^{-\frac{1}{6}}]}\,=\,e^{60} 
\nonumber \\
&&\Rightarrow~~~~~
b_1a_2^{-\frac{1}{6}}\times\,10^6\,sec^{-\frac{1}{6}}\,=\,60 
\nonumber \\
&&
\Rightarrow\,~~~~
b_1a_2^{-\frac{1}{6}}\,
\simeq\,6\times\,10^{-5}\,sec^{\frac{1}{6}}
~~~~
\Rightarrow\,~~~~\ln{\beta}\,=\,8.8\times\,10^{-8}
\label{a16b}
\end{eqnarray}
If we assume 60 e-fold expansion and $t_{is}= \,=\,10^{-28}\,sec$, 
$t_{il}\,=\,t_{is}+10^{-34}\,sec$
\begin{eqnarray}
\frac{a(t_{is})}{a(t_{il})}&\simeq&
\exp{\{-b_1\,
(a_2)^{-\frac{1}{6}}
(1\times\,10^{-30})^{-\frac{1}{6}}[(100)^{-\frac{1}{6}}\,-\,
(100.0001)^{-\frac{1}{6}}]\}}
\,=\,e^{60} \nonumber \\
&&\Rightarrow~~~~~
b_1a_2^{-\frac{1}{6}}\times\,7.7\times\,10^{-3}
\,=\,60 
\nonumber \\
&&\Rightarrow\,~~~~b_1a_2^{-\frac{1}{6}}\,\simeq\,7.8\,\times\,10^3
~~~~
\Rightarrow\,~~~~\ln{\beta}\,=\,11
\label{a16d}
\end{eqnarray}
Note that
\begin{equation}
b_1a_2^{-\frac{1}{6}}\,=\,
b_1(a_2t_0)^{-\frac{1}{6}}
t_0^{\frac{1}{6}}\,=\,
t_0^{\frac{1}{6}}\,\ln{\beta}
\label{a16g}
\end{equation}
Some other values of $t_{is}$, $t_{il}$ and $\ln{\beta}$ for 60 e-fold 
expansion are 
\begin{eqnarray}
&&t_{is}\,=\,10^{-30}\,sec     
~~~~~~
t_{il}\,=\,t_{is}+10^{-34}
~~~~\Rightarrow~~~~\ln{\beta}\,=\,
5.57\times\,10^{-2}
\label{f8a} \\
&&t_{is}\,=\,10^{-24}\,sec     
~~~~~~
t_{il}\,=\,t_{is}+10^{-34}\,sec
~~~~\Rightarrow~~~~\ln{\beta}\,=\,
5.57\times\,10^{5}
\label{f8b} \\
&&t_{is}\,=\,
10^{-28}\,sec\,+\,     
10^{-34}\,sec     
~~~~~~
t_{il}\,=\,t_{is}+2\times\,10^{-34}\,sec
~~~~\Rightarrow~~~~\ln{\beta}\,=\,
5.72
\label{f8c} \\
&&t_{is}\,=\,
10^{-28}\,sec\,+\,     
10^{-34}\,sec     
~~~~~~
t_{il}\,=\,t_{is}+2.4\times\,10^{-34}\,sec
~~~~\Rightarrow~~~~\ln{\beta}\,=\,
8.15
\label{f8d} \\
&&t_{is}\,=\,
10^{-29}\,sec\,
~~~~~~
t_{il}\,=\,t_{is}+5\times\,10^{-34}\,sec
~~~~\Rightarrow~~~~\ln{\beta}\,=\,
0.13
\label{f8e} 
\end{eqnarray}
One notices that (\ref{a16b}), 
(\ref{f8a}), (\ref{f8c}), (\ref{f8e}) 
are consistent with (\ref{f1}) while the others are not. However it seems 
that the values in Table I exclude the values of $\ln{\beta}$ much smaller 
than 1. This excludes the options in (\ref{a16b}) and (\ref{f8a}) as well. 
Hence 
the viable values seem to be (\ref{f8c}) and (\ref{f8e}) and all values of 
parameters between them and close to these values. This offers a wide 
range of $t_{is}$ between $10^{-28}\,sec$ and $10^{-29}\,sec$.   
It is evident that all phenomenologically viable values may be obtained 
by adjusting 
$t_{is}$ in the  
$t_{is}=10^{-29}\,sec$ -- 
$t_{is}=10^{-28}\,sec$ 
range that corresponds to a lower scale inflation \cite{lower}.

A comment is in order at this point. From Eq.(\ref{a4}) we see that just 
at the end of the inflationary era
\begin{eqnarray}
&&b_1(p_1+p_2a_2t_{mi})+(a_2t_{mi})^{\frac{1}{6}}(-7p_1+5p_2a_2t_{mi})\,\leq
\,0 \label{a17a}\\
&&\Rightarrow~~~~
p_1\frac{b_1}{(a_2t_{mi})^{\frac{1}{6}}}-7p_1+\epsilon=0 \label{a17b} \\
&&\left(\gamma_{mi}^{-\frac{1}{6}}\ln{\beta}-7\right)
(1-\xi+\frac{1}{6}\ln{\beta})
+\epsilon^\prime\,=\,0 \label{a17c} \\
&&~~~~\epsilon=\frac{b_1}{(a_2t)^{\frac{1}{6}}}p_2a_2t+5p_2a_2t~,~~
\epsilon^\prime=\frac{\epsilon}{\beta}=\gamma_{mi}
(\gamma_{mi}^{-\frac{1}{6}}\ln{\beta}+5)(\xi-\frac{1}{6}\ln{\beta})
\nonumber
\end{eqnarray}
where the first two terms in (\ref{a17b}) are the dominant terms and 
$\epsilon$ (and $\epsilon^\prime$) is small with respect to the others. 
The fact that $\epsilon^\prime$ in and at the end of inflationary era is 
small implies that either 
$(\gamma_{mi}^{-\frac{1}{6}}\ln{\beta}-7)$ 
or $(1-\xi+\frac{1}{6}\ln{\beta})$ is small. Taking 
$t_{mi}\sim\,t_{il}\sim\,10^{-28}--10^{-29}$ i.e. 
$\gamma_{mi}=\frac{t_{mi}}{t_0}\sim\,10^{-46}$ implies that  
$(\gamma_{mi}^{-\frac{1}{6}}\ln{\beta}-7)$ is not small unless 
$\ln{\beta}$ is extremely small. Therefore 
$(1-\xi+\frac{1}{6}\ln{\beta})$ should be small if deceleration era 
starts just after the inflationary era. This may be provided by 
taking $\xi$ a little bit larger than 1 and $\ln{\beta}$ small. For 
example one may take $\xi=1.05$ and $\ln{\beta}\sim\,0.3$ (i.e. 
$\beta\sim\,1.35$). Otherwise one should take the start of the 
deceleration era much later than the standard inflationary era (i.e. 
the inflationary era is much longer than the standard inflationary 
times). Although this option seems to be a less acceptable option it is, 
in fact, the more reasonable choice. This is due to the fact that we 
neglect radiation in this study. In the realistic case there is a 
radiation dominated era just after the inflationary era. Radiation like 
matter drives the universe towards deceleration. Therefore if we add 
radiation to the model it is effect will be an earlier start of 
deceleration era compared to the radiationless case. This explains why the 
time of the start of the deceleration period almost coincides with the 
time of start of the matter dominated era in Table (\ref{table-tmi-tmf}) 
unless $\beta$ is extremely close to 1. In other words the values 
of parameters become less reliable as we get closer to the radiation 
dominated era. We should keep this in mind as we analyze the 
observational data.

Now we apply the values obtained to the time of reionization, $t_{ri}$. In 
fact we expect, at most, a rough agreement with data since $t_{ri}$ goes 
deeper 
into the matter dominated era where neglecting  baryonic 
matter becomes more questionable.
\begin{equation}
\frac{a_0}{a(t_{ri})}
\,=\,\frac{\exp{[b_1(a_2t_{ri})^{-\frac{1}{6}}]}}{(p_1+p_2a_2t_{ri})}
\,=\,\frac{\beta^{\gamma_{ri}^{-\frac{1}{6}}}}
{(1-\xi)\beta 
+\frac{1}{6}\ln{\beta}+(\xi\beta-\frac{1}{6}\ln{\beta})\gamma_{ri}}
\label{a231}
\end{equation}
where $\gamma_{ri}=\frac{t_{ri}}{t_0}$.
The observational value of 
$\frac{a_0}{a(t_{ri})}$ is 
$12\pm\,1.4$ and the corresponding $\Lambda$CDM value of $t_{ri}$ is
430$^{+90}_{-70}$ Myr that corresponds to $\gamma_{ri}$ in the interval 
$0.0225~---~0.0472$ if one assumes a loose bound on the value of $t_0$, 
$t_0\,=\,11~---~16\;Gyrs$ in the light of the values of $t_0$ from 
different observations mentioned before. One may plot 
$\frac{a_0}{a(t_{ri})}$ versus $\gamma=\frac{t}{t_0}$ for the 
phenomenologically 
relevant values of $\xi$ and various $\beta$ 
values. 
The allowed 
intervals of $\gamma_d$ for $\frac{a_0}{a(t_{ri})}$ in the interval 10.6 - 
12.4 for various values of $\xi$ and $\beta$ may be 
found in Table \ref{table-atd-g}. 
\begin{table}[h] 
\begin{tabular}{|c|c|c||c|c|c|} \hline
$\xi$&$\beta$&$\gamma_{ri}$&
$\xi$&$\beta$&$\gamma_{ri}$\\
\hline 
0.8&2&0.0016~---~0.0023&
0.8&5&
0.0115~---~0.016 \\
\hline 
0.8&10&0.023~---~0.03&
0.8&20&0.037
~---~0.046 \\
\hline 
0.8&50&0.058~---~0.068&
0.8&100&0.074~---~0.086\\
\hline 
0.8&200&0.091~---~0.105&
0.8&500&0.116~---~0.128 \\
\hline 
0.8&1000&0.13~---~0.146&
0.8&2000&0.15~---~0.162\\
\hline 
1&2&0.028~---~0.041&
1&5&
0.029~---~0.041\\
\hline 
1&10&0.04~---~0.051&
1&20&0.054
~---~0.066 
\\
\hline 
1&50&0.079~---~0.086&
1&100&0.089~---~0.104
\\
\hline 
1&200&0.105~---~0.12&
1&500&0.13~---~0.142 \\
\hline 
1&1000&0.142~---~0.16&
1&2000&0.16~---~0.18\\
\hline 
1.2&2&0.17~---~0.19&
1.2&5&
0.105~---~0.125\\
\hline 
1.2&10&0.09~---~0.105&
1.2&20&0.091
~---~0.104 \\
\hline 
1.2&50&0.098~---~0.116&
1.2&100&0.114~---~0.127\\
\hline 
1.2&200&0.125~---~0.142&
1.2&500&0.144~---~0.162\\
\hline 
1.2&1000&0.159~---~0.177&
1.2&2000&0.173~---~0.192\\
\hline 
\end{tabular}
\caption[b]{The allowed range of values of 
$\gamma_{ri}=\frac{t_{ri}}{t_0}$ for 
various values of
$\xi$, $\gamma$ with $\frac{a_0}{a(t_d)}$ in the range $\,10.6\,--\, 
12.4$}  
\label{table-atd-g}
\end{table}
It seems that the values of $\beta$ compatible with data are 10, 20 for 
$\xi$= 0.8; 2, 5, 10 for  $\xi$= 1; and none for $\xi$=1.2. However one 
should keep in mind that a more detailed analysis may give a wider range 
of parameters since the age calculations in the data 
analysis \cite{Komatsu2009} use a restricted form for dark energy where 
Hubble constant may be expressed in terms of density parameters where 
matter is 
assumed dust-like and a restricted form of variation of dark energy with 
redshift, and a restricted class of equations of state for dark energy 
where dark energy is not  entangled with matter as pointed out before.
Therefore reanalysis of data in the context of this model is necessary to 
reach a more precise and more definite conclusion.
Another factor for poorer agreement with data is that we neglect the 
contribution of baryonic matter whose contribution in matter 
dominated period is greater.

Next consider the data for the time of decoupling and the corresponding 
redshift; $z_*\,\simeq\,1090$ 
\begin{equation} \frac{a_0}{a(t_*)} 
\,=\,\frac{\exp{[b_1(a_2t_*)^{-\frac{1}{6}}]}}{(p_1+p_2a_2t_*)} 
\,=\,\frac{\beta^{\gamma_*^{-\frac{1}{6}}}} {(1-\xi)\beta 
+\frac{1}{6}\ln{\beta}+(\xi\beta-\frac{1}{6}\ln{\beta})\gamma_*} 
\label{a231} \end{equation} 
where $\gamma_*=\frac{t_*}{t_0}$. One may plot 
$\frac{a_0}{a(t_*)}$ 
versus $\gamma=\frac{t}{t_0}$ for various values of 
$\xi$ and $\beta$. The values of $\gamma_*$ corresponding to the 
observational value of 
$\frac{a_0}{a(t_*)}\sim\,1090$ are given in Table \ref{table-at-dc}.
\begin{table}[h] 
\begin{tabular}{|c|c|c||c|c|c|} \hline
$\xi$&$\beta$&$\gamma_{ri}$& $\xi$&$\beta$&$\gamma_{ri}$\\
\hline 
0.8&2&$1.42\times\,10^{-6}$&
0.8&5&
$7.45\times\,10^{-5}$
\\
\hline 
0.8&10&
$3.3\times\,10^{-4}$&
0.8&20&
$9.05\times\,10^{-4}$\\
\hline 
0.8&50&
$2.32\times\,10^{-3}$&
0.8&100&
$3.99\times\,10^{-3}$\\
\hline 
0.8&200&
$6.17\times\,10^{-3}$&
0.8&500&
$9.86\times\,10^{-3}$
\\
\hline 
0.8&1000&
$1.32\times\,10^{-2}$&
0.8&2000&
$1.71\times\,10^{-2}$
\\
\hline 
1&2&
$3.9\times\,10^{-6}$&
1&5&
$1.15\times\,10^{-4}$
\\
\hline 
1&10&
$4.44\times\,10^{-4}$&
1&20&
$1.12\times\,10^{-3}$
\\
\hline 
1&50&
$2.7\times\,10^{-3}$&
1&100&
$4.5\times\,10^{-3}$
\\
\hline 
1&200&
$6.8\times\,10^{-3}$&
1&500&
$1.06\times\,10^{-2}$
\\
\hline 
1&1000&
$1.425\times\,10^{-2}$&
1&2000&
$1.81\times\,10^{-2}$
\\
\hline
1.2&2&
$7.91\times\,10^{-2}$&
1.2&5&
$7.45\times\,10^{-5}$
\\
\hline 
1.2&10&
$3.3\times\,10^{-4}$&
1.2&20&
$1.56\times\,10^{-3}$
\\
\hline 
1.2&50&
$3.34\times\,10^{-3}$&
1.2&100&
$5.28\times\,10^{-3}$
\\
\hline 
1.2&200&
$7.7\times\,10^{-3}$&
1.2&500&
$1.17\times\,10^{-2}$
\\
\hline 
1.2&1000&
$1.53\times\,10^{-2}$&
1.2&2000&
$1.95\times\,10^{-2}$ \\
\hline 
\end{tabular}
\caption[b]{The allowed range of values of 
$\gamma_{*}=\frac{t_{*}}{t_0}$ for 
various values of
$\xi$, $\gamma$ with $\frac{a_0}{a(t_d)}\,\sim\,1090$}  
\label{table-at-dc}
\end{table}
Inspection of the table suggests there are no of the values are $\xi$ and 
$\beta$ compatible 
with observational value
$t_*\,\simeq\,3.8\times10^5\,yr$ \cite{Komatsu2009} that corresponds to 
the interval 
$\gamma_*=\frac{t_*}{t_0}\,= \,
2.375\times10^{-5}~---~3.4545\times10^{-5}$ (provided that 
$t_0\,=\,(11~---~16)\,Gyrs$) except for $\xi=0.8,\,1 $ and $\beta$ 
somewhere between 2 and 5 (i.e. $\sim$ 3.7, 3) while the values for 
$\xi=0.8$, $\beta=5$ and $\xi=1.2$, $\beta=5$ are close to the relevant 
values. In fact this poor agreement with those given in \cite{Komatsu2009} 
is expected. In addition to the reasons mentioned for the reionization 
time there is an important additional source of discrepancy. The time of 
matter radiation decoupling time is quite close to the radiation dominated 
era. The ratio of radiation in this period in the order of a fourth of the 
total energy density at this time while this model neglects the 
contribution of radiation.

To summarize the results of this section can be stated as follows: We 
have seen that the 
predictions of this model for each of 
$\left(\frac{\ddot{a}}{a}/G_{00}\right)_{t=t_0}$,  
$\left(\frac{G_{11}}{g_{11}}/G_{00}\right)_{t=t_0}$, 
$a_0/a(t_d)$,  
$a_0/a(t_{ri})$ are compatible with observations although not with central 
values given in literature. The prediction of the model for $a_0/a(t_*)$ 
is partially consistent with observational values. In fact the relatively 
less compatibility for the decoupling time $t_*$ is 
expected since the radiation-matter decoupling time is close to the 
radiation dominated era while radiation is ignored in this  study. 
I have also shown that an inflationary era naturally fits the model.
One may  
consider the simultaneous compatibility of the predictions of all these 
parameters with observations as well.  In all cases there is wide range of 
$\beta$'s compatible with Eq.(\ref{f1}). 
The values of $\beta$ allowed by 
(\ref{c8}) includes the values allowed by 
Table \ref{table-G00-ac-present}, 
that is, 
Table \ref{table-G00-ac-present} and (\ref{c8}) are compatible while 
Table \ref{table-G00-ac-present} is more restrictive. The values of 
$\gamma_d$ in Table \ref{table-transition-acc-dec} that are compatible 
with Table \ref{table-atd-b} are 
$\xi$=0.8 $\Rightarrow$ $\gamma_d$=0.4 or $\gamma_d$=0.5,  
$\xi$=1 $\Rightarrow$ $\gamma_d$=0.5 or $\gamma_d$=0.8, 
$\xi$=1.2 $\Rightarrow$ hardly $\gamma_d$=0.8. The values of $\xi$, 
$\gamma_d$ in Table \ref{table-atd-b} whose $\beta$ values compatible with 
the $\beta$ values in  
Table \ref{table-G00-ac-present} are 
$\xi$=0.8, $\gamma_d$=0.4, $\gamma_d$=0.5, 
$\beta$= 2.14 -- 12.2;   
$\xi$=1, $\gamma_d$=0.5 $\Rightarrow$ $\beta$= 4 -- 144;   
$\xi$=1, $\gamma_d$=0.6 $\Rightarrow$ $\beta$= 130 -- 147;   
$\xi$=1.2, $\gamma_d$=0.6 $\Rightarrow$ 
Table \ref{table-G00-ac-present} and
Table \ref{table-atd-b} are compatible except for lowest 
values of $\beta$. Mostly  
Table \ref{table-G00-ac-present} is more restrictive than
Table \ref{table-atd-b}. 
The values of $\gamma_{ri}$ in Table \ref{table-atd-g} 
that are more compatible with observational value $\gamma_{ri}$= 0.0225 
-- 
0.0472 in literature 
\cite{Komatsu2009} seem to prefer $\beta$ in the range 2 -- 10. I do 
not use Table \ref{table-at-dc} to constraint $\beta$ since the time of 
decoupling is close to the radiation dominated era while we do ignore 
radiation, so the reliability of the values obtained is questionable, and 
an order of magnitude compatibility is enough.  We see 
that compatibility of the values of all these tables seem to prefer values 
$\xi$ = 0.8 - 1, $\beta$ = 2 - 10. However these values of $\beta$ are at 
the edge of the observationally allowed values rather than being centrally 
allowed values.  
The 
limited overall compatibility of the results of this model with 
observations may 
either be due to this model being simply a toy model or the 
inapplicability of some of the assumptions of the analysis in literature 
to this model such as Eq.(\ref{f4}) and 
$\omega(a)\,=\,-\omega_0+\omega_1(1-a)$ 
or a combination of both. In fact, even a standard analysis may be 
enough to check the viability of this model beyond a toy model for small 
enough redshift bins. For example, it seems that the 
allowed value of the equation of state for dark energy, $\omega_{DE}$ 
at the smallest redshift bin in Figure 14 of \cite{Amanullah} 
may be as large as -1/3 while the $G_{11}/G_{00}$ 
versus $\beta$ graph for this model at present time (for $\xi=1$) gives 
$G_{11}/G_{00}\simeq\,-0.45$ (that corresponds to $\omega_{DE}\sim\,-0.65$ 
for $\Omega_{DE}=0.74$) for $\beta\sim\,3$. A definite conclusion needs a 
detailed comprehensive reanalysis of all data in the light of this model 
in a separate study.

\section{conclusion}
I have considered a model where inflationary era and (dark) matter 
dominated eras are induced by a scalar field $\phi_1$ while the dark 
energy dominated era is induced by another scalar $\phi_2$. These fields 
may be either considered to be fundamental fields or as effective 
classical fields. I prefer to consider them as classical fields rather 
than true fundamental fields. 
I have neglected the effects of baryonic matter and radiation. A rough 
phenomenological analysis 
of cosmic data gives an order of magnitude agreement with data. This is 
encouraging for future studies in this direction. One must include 
baryonic 
matter and radiation to obtain a more realistic model. However this is not 
an easy task. 
First difficulty is that baryonic matter and radiation should be included 
after the time of inflation because it should be produced by the decay of 
one of the scalars (probably by  $\phi_1$). Second, even when one includes 
them in ad hoc way this modifies the metric. Hence one must find the scale 
factor that corresponds to inclusion of the baryonic matter and radiation 
and this not a trivial task. Another point that needs further study is a 
more detailed and comprehensive analysis of the available parameter space 
and to find the most optimal set. Yet another point for further study is 
the study of cosmological perturbations produced in the inflationary epoch. 
The inflation obtained here is a standard slow-roll inflation 
with the canonical kinetic terms for the scalars. Therefore the general 
form of the perturbations is the same as the usual slow-roll case 
\cite{Weinberg}. However a detailed study of the perturbations in 
this model should be obtained to compare with the expectations of the 
other models for data 
to be obtained in future cosmological observations. All these points need 
further separate studies.

\begin{acknowledgments}
I would like to thank Professor Joan Sol{\`{a}}, and Professor Eduardo I.
Guendelman for reading the manuscript and for their valuable comments.
\end{acknowledgments}



\bibliographystyle{plain}

\end{document}